\magnification=1200
\def\ump{1-p_0/\k}
\def\d{\delta}

\def\k{\kappa}\def\l{\lambda}\def\m{\mu}\def\n{\nu}\def
\p{\pi}\def\t{\tau}
\def\y{\eta}\def\x{\xi}

\def\mn{{\mu\nu}}

\def\fe{field equations }

\def\pb{Poisson brackets }

\def\poi{Poincar\'e }

\def\SR{special relativity }

\def\lt{Lorentz transformations }
\def\tl{transformation law }

\def\section#1{\bigskip\noindent{\bf#1}\smallskip}
\def\nota{\footnote{$^\dagger$}}

\def\PL#1{Phys.\ Lett.\ {\bf#1}}
\def\PRL#1{Phys.\ Rev.\ Lett.\ {\bf#1}}
\def\PR#1{Phys.\ Rev.\ {\bf#1}}\def\CQG#1{Class.\ Quantum Grav.\ {\bf#1}}

 \def\IJMP#1{Int.\ J. Mod.\ Phys.\ {\bf #1}}
\def\MPL#1{Mod.\ Phys.\ Lett.\ {\bf #1}} 

\def\AoP#1{Ann.\ Phys.\ {\bf#1}}
\def\hep#1{{\tt hep-th/#1}}

\def\ref#1{\medskip\everypar={\hangindent 2\parindent}#1}
\def\beginref{\begingroup
\bigskip
\centerline{\bf References}
\nobreak\noindent}
\def\endref{\par\endgroup}

{\nopagenumbers
\line{}
\vskip40pt
\centerline{\bf The Magueijo-Smolin model of Deformed Special Relativity}
\centerline{\bf from five dimensions}
\vskip60pt
\centerline{{\bf S. Mignemi}\nota{e-mail: smignemi@unica.it}}
\vskip10pt
\centerline {Dipartimento di Matematica, Universit\`a di Cagliari}
\centerline{viale Merello 92, 09123 Cagliari, Italy}
\smallskip
\centerline{and INFN, Sezione di Cagliari}

\vskip100pt
\centerline{\bf Abstract}
\vskip10pt
{\noindent
It is known that
the space of momenta of DSR can be identified with de Sitter space.
In this paper, we discuss the relation of the noncanonical phase space
of the Magueijo-Smolin model of DSR with the canonical 5-dimensional
phase space in which the de Sitter space of momenta is embedded.
We suggest that in analogy with the momentum variables, also
the position variables should be constrained to lie on a null
hypersurface of the five-dimensional space.}
\vskip100pt
\vfil\eject}

The purpose of doubly special relativity (DSR) theories is to
give an effective description of quantum gravity effects on particle
dynamics at energies near the Planck scale, by postulating a nonlinear
(deformed) action of the Lorentz group on momentum space,
such that the Planck energy $\k$ is left invariant [1-4].

Since the deformation is not uniquely defined, one can obtain many
different realizations of the theory.
An advance in the understanding of this problem was given
by the observation
that the momentum space of DSR models can be thought as
a four-dimensional hyperboloid embedded in a five dimensional target
space [5].
The choice of coordinates on the hyperboloid corresponds to different
DSR models.

Although this picture is very suggestive, no convincing interpretation
has been advanced for the dual five-dimensional position space. It must
be noticed however that already in a four-dimensional setting
the realization of position space in DSR theories is ambiguous, since
it is not determined in a natural way by the DSR postulates
(see e.g.\ [6] and references therein).
For example, although a realization in terms of noncommutative coordinates
appears more natural, it is also possible to adopt standard noncommuting
coordinates [10].

Moreover, the physical meaning of a fifth position coordinate is unclear.
In [7] a proposal was advanced based on the existence of a linear
realization of the deformed Lorentz group in five dimensional momentum space.
The fifth position coordinate was identified with the evolution parameter of
the field equations in a commutative spacetime.
A different interpretation based on the hamiltonian formalism was put
forward in [8]. In that case the fifth coordinate is related to an
arbitrary choice of gauge, but its relation with the physical observables
is unclear.

In this paper we use some recent results [9] to identify the correspondence
between the physical coordinates of the MS model [4], which is probably the
simplest realization of DSR from an algebraic point of view, and the
5-dimensional target space variables of [5].
Following [7], we also identify a formal fifth spacetime coordinate with
the invariant evolution parameter which appears in the field equations.
In this way we constrain the 5-dimensional position coordinates to form
a null vector, so that only four coordinates are independent, as required
by physics.
\bigskip

We use the following conventions: $A=0,\dots,4$ are target space indices,
$\m=0,\dots,3$ are spacetime indices, $i=1,\dots,3$ are spatial indices.
We always use lower indices, which are
summed by means of the flat metric $\y_\mn={\rm diag}\ (1,-1,-1,-1)$.
The modulus of a 4-vector $A_\m$ is denoted by $A\equiv\sqrt{A_\m A_\m}$.

The MS model is defined by nontrivial transformation laws of the physical
momenta $p_\m$ under boosts [4]. For infinitesimal boosts in the $i$-th
direction, the momenta transform as
$$\d_ip_0=(\ump)p_i,\qquad\d_ip_j=\d_{ij}p_0-p_ip_j/\k,\eqno(1)$$
while they transform in the standard way under rotations.
The quantity
$${p^2\over(\ump)^2}\eqno(2)$$
is invariant under the deformed transformations (1).

The MS model does not specify the transformation law under boosts for the
position coordinates $x_\m$. The most natural choice is based on the
requirement that they transform covariantly with respect to the momenta [10]:
$$\d_ix_0=x_i+p_ix_0/\k,\qquad\d_ix_j=\d_{ij}x_0+p_ix_j/\k.\eqno(3)$$
This can be achieved if the phase space coordinates satisfy the \pb [10,11]
$$\eqalignno{&\{x_0,x_i\}=x_i/\k,\quad\{p_0,p_i\}=0,\quad\{x_0,p_0\}=\ump,&\cr
&\{x_i,p_j\}=-\d_{ij},\quad\{x_0,p_i\}=-p_i/\k,\qquad\{x_i,p_0\}=0,&(4)}$$
which are typical of DSR models.
Notice that one is forced to use noncommutative coordinates.
This definition also yields the classical \tl for the velocity and fixes the
speed of light as the limit velocity [10,11]. The main peculiarity of (3)
is the momentum dependence of the transformations of the spacetime coordinates
under boosts.

In a recent paper [9], it was shown that one can define auxiliary variables
\footnote*{This result had been independently anticipated in [11].}
$$P_\m={p_\m\over\ump},\qquad X_\m=\left(\ump\right)x_\m,\eqno(5)$$
that satisfy canonical \pb
$$\{X_\m,P_\n\}=\y_\mn,\qquad\{X_\m,X_\n\}=\{P_\m,P_\n\}=0.\eqno(6)$$
These coordinates are unphysical, but are helpful in order to convert the
results
of \SR to those of the MS model. In particular, the physical quantities have
the standard expression in terms of the auxiliary coordinates $X_\m$, $P_\m$.
To derive their expression in the MS model it is then sufficient to write
them in terms of the physical coordinates $x_\m$, $p_\m$.

For example, the generators $J_\mn$ of the Lorentz transformations take the
form $J_\mn=X_\m P_\n-X_\n P_\m$. Using (5), one easily deduces that the
Lorentz generators have the standard form $J_\mn=x_\m p_\n-x_\n p_\m$ also in
terms of the MS coordinates [9].
From the \pb (4) one can then recover the deformed \lt for $x_\m$ and $p_\m$.

\bigbreak
The purpose of this letter is to relate the results of [9] to the
five-dimensional formalism of [5] and to give an interpretation of the
fifth coordinate as attempted in [7,8].

In [5] it was shown that one can identify the space of momenta with an
hyperboloid of equation $\p^2-\p_4^2=-\k^2$ in a 5-dimensional momentum space
of coordinates $\p_A$ and metric $\y_{AB}=(1,-1,-1,-1,-1)$.
The choice of different coordinates $\tilde P_\m$ on the hyperboloid
gives rise to different realizations of DSR. For the Snyder basis, for example,
$$\tilde P_\m={\k\over\p_4}\,\p_\m.\eqno(7)$$
Introducing the 5-dimensional position variables $\x_A$ canonically conjugate
to $\p_A$, one can write down the variables $\tilde X_\m$ canonically
conjugated to
$\tilde P_\m$ as
$$\tilde X_\m={\p_4\over\k}\,\x_\m.\eqno(8)$$

We identify the variables $\tilde P_\m$ and $\tilde X_\m$ with $P_\m$ and
$X_\m$ defined above.
From (5) and (7)-(8) then follows
$$p_\m={\k\over\p_0+\p_4}\, \p_\m,\qquad
x_\m={\p_0+\p_4\over\k}\, \x_\m.\eqno(9)$$

At this point one might apply the hamiltonian formalism introduced in [8] for
the Snyder model.
Denoting with a bar the variables of [8], they can be written in terms
of $X_A$ and $P_A$ as $\bar p_\m=P_\m$, $\bar x'_\m=X_\m-X_AP_A\,P_\m/\k^2$,
$\bar T=X_AP_A$.
Then our variables $X_\m$ and $P_\m$ are analogous to $y_\m$ and $q_\m$ in
eq.\ (30) of ref. [8].

Instead, we prefer to introduce a different interpretation of 5-dimensional
spacetime, closer to that proposed in [7]. In particular, we wish to
constrain  the 5-dimensional coordinates so that the fifth coordinate
coincides with the evolution time $\t$ that parametrizes the trajectories of
the particles. The relation of $\t$ with the spacetime coordinates, first
derived in [11], has often been overlooked, but is important, at least in a
lagrangian formalism.
Since it is by definition invariant under the deformed Lorentz transformations,
$d\t^2$ can be identified with the line element of the 4-dimensional
spacetime [10,11].
Although in this context the introduction of the coordinate $\t$ is purely
formal, it gives a more elegant formulation of the action principle.

From (7) we may define the fifth component of $P_\m$ as $P_4=\k$. We
also define the fifth component of $X_\m$ by imposing that $X_A$ be a null
vector, i.e.\ $X_4^2=X^2$. Then,
$$X_4=X={\p_4\over\k}\,\x.\eqno(10)$$
It is easy to check that $X_4$ is left invariant by the deformed \lt generated
by $J_\mn$.
By the previous definition, $X_4$ coincides with the invariant affine parameter
$\t$ which parametrizes the trajectories of point particles [10,11]. In analogy
with [7], we consider therefore a 5-dimensional spacetime with coordinates
$(x_\m,\t)$.
One may interpret this choice as constraining the motion to the hypersurface
$\x_A^2=0$, in analogy with the constraint $\p_A^2=-\k^2$ on momentum
space. This appears to be the most natural condition to be imposed in order to
reduce to four the number of independent position coordinates.

From
$$\{X_\m,X_4\}=0,\qquad\{P_\m,X_4\}={X_\m\over X_4},\eqno(11)$$
follows
$$\{x_\m,\t\}=(\ump)^2\ {x_0x_\m/\k\over\t},\qquad
\{p_\m,\t\}=(\ump)^2\ {x_\m-x_0p_\m/\k\over\t}.$$

The dynamics of a free particle of mass $m$ can be obtained by varying
the action
$$\eqalignno{I=&\int d\t\left[\dot X_\m P_\m-{\l\over2}(P^2-m^2)\right]&\cr
=&\int d\t\left[{p_\m\over\ump}\ {d\over d\t}\big[(\ump)x_\m\big]-{\l\over2}
\left({p^2\over(\ump)^2}-m^2\right)\right],&(12)}$$
with $\t$ defined above. This is in contrast with other formulations
where $\t$ is an external parameter whose properties are not specified.
The action is equivalent up to total derivatives to those given in refs.\
[9,10], where the \fe following from (12) are discussed.
\bigskip

To conclude, we consider the extension from the Lorentz to the \poi algebra.
This is achieved by adding the translations generators $T_\m$ to the Lorentz
algebra. It must be noticed, however, that the generators $T_\m$ are not
determined
uniquely. Usually they are identified (at least implicitly) with the momentum
coordinates $p_\m$. With this definition, they act linearly on the space of
momenta, but the \poi algebra is deformed. In particular,
the boost generators $N_i$ have nonlinear \pb with the translation generators,
$$\{N_i,T_0\}=T_i-T_0T_i/\k,\qquad\{N_i,T_j\}=\d_{ij}T_0-T_iT_j/\k.\eqno(13)$$
The action of the translations on the coordinates following from this definition
is
$$\eqalignno{&\{T_0,x_0\}=-(\ump)\qquad\{T_0,x_i\}=0,&\cr
&\{T_i,x_0\}=p_i/\k,\qquad\{T_i,x_j\}=\d_{ij}.&(14)}$$

An alternative possibility [9] is to define $T_\m=P_\m$. In this case the
standard form of the \poi algebra is preserved, but the translation operator
acts nontrivially on the momenta.
The action of the translations on the coordinates takes a neater form,
$$\eqalignno{&\{T_0,x_0\}={-1\over\ump}\qquad\{T_0,x_i\}=0,&\cr
&\{T_i,x_0\}=0,\qquad\{T_i,x_j\}={\d_{ij}\over\ump}.&(15)}$$
Its effect is a sort of momentum-dependent dilation of time and lengths under
translations.

At the classical level, the difference between the two representations of
translations is not great, since in both cases $\{T_\m,p_\n\}=0$, but at the
quantum level is more evident. In particular, the natural law of addition of
momenta, arising from translation invariance, is linear in the first case,
while in the second case coincides with that proposed in [12], which implies
the existence of a maximum energy $\k$. The most natural choice in the context
of DSR for both the addition law of momenta and the transformation of coordinates
seems to be that determined by $T_\m=P_\m$.
However, since the one-particle action (12) is invariant under the action of
any generator $T_\m$ which is function only of $p_\m$, at this stage one could
choose $T_\m$ arbitrarily. This posibility may be of help in the solution of the
so-called soccer ball problem. In any case, the correct addition law can be derived
only after the general form of the interaction between particles has been established.

\vfill\eject

\beginref
\ref [1] G. Amelino-Camelia, \IJMP{D11}, 35 (2002), \PL{B510}, 255 (2001).
\ref [2] J. Lukierski, H. Ruegg and W.J. Zakrzewski, \AoP{243}, 90 (1995).
\ref [3] J. Kowalski-Glikman, \MPL{A17}, 1 (2002).
\ref [4] J. Magueijo and L. Smolin, \PRL{88}, 190403 (2002).
\ref [5] J. Kowalski-Glikman, \PL{B547}, 291 (2002);
J. Kowalski-Glikman, S. Nowak, \CQG{20}, 4799 (2003).
\ref [6] P. Gal\'an, G. Mena Marugan, \IJMP{D16}, 1133 (2007).
\ref [7] A.A. Deriglazov, B.F. Rizzuti, \PR{D71}, 123515 (2005).
\ref [8] F. Girelli, T. Konopka, J. Kowalski-Glikman, E.R. Livine, \PR{D73}, 045009 (2006).
\ref [9] S. Ghosh, P. Pal, \PR{D75}, 105021 (2007).
\ref [10] S. Mignemi, \PR{D68}, 065029 (2005).
\ref [11] A. Granik, \hep{0207113}.
\ref [12] S. Judes and  M. Visser, \PR{D68}, 045001 (2003).

\endref
\end